\documentclass{emulateapj}

\newcommand{\lapprox }{{\lower0.8ex\hbox{$\buildrel <\over\sim$}}}
\newcommand{\gapprox }{{\lower0.8ex\hbox{$\buildrel >\over\sim$}}}
\def\amin{\ifmmode^{\prime}\else$^{\prime}$\fi}
\def\asec{\ifmmode^{\prime\prime}\else$^{\prime\prime}$\fi}
\def\ctss{$10^{-2}$ count s$^{-1}$}
\def\ergcms{erg~cm$^{-2}$~s$^{-1}$}
\def\ergs{erg~s$^{-1}$}
\def\ROSAT{\it ROSAT}

\shorttitle{NO NEW INSs IN SDSS DR4}
\shortauthors{AG{\" U}EROS ET AL.}

\begin{document}

\title{No Confirmed New Isolated Neutron Stars In The SDSS Data Release 4\altaffilmark{1}}

\author{
Marcel A.\ Ag\"ueros\altaffilmark{2,7},
Bettina Posselt\altaffilmark{3,8},
Scott F.\ Anderson\altaffilmark{4},
Philip Rosenfield\altaffilmark{4},
Frank Haberl\altaffilmark{5},
Lee Homer\altaffilmark{4},
Bruce Margon\altaffilmark{6},
Emily R.\ Newsom\altaffilmark{2},
Wolfgang Voges\altaffilmark{5}
}

\altaffiltext{1}{Based on observations obtained at the Gemini Observatory, which is operated by the Association of Universities for Research in Astronomy (AURA) under a cooperative agreement with the NSF on behalf of the Gemini partnership: the National Science Foundation (United States), the Science and Technology Facilities Council (United Kingdom), the National Research Council (Canada), CONICYT (Chile), the Australian Research Council (Australia), CNPq (Brazil) and CONICET (Argentina). Also includes observations obtained with the Apache Point Observatory $3.5$-m telescope, which is owned and operated by the Astrophysical Research Consortium.}
\altaffiltext{2}{Columbia University, Department of Astronomy, 550 West 120th Street, New York, NY 10027, USA; marcel@astro.columbia.edu}
\altaffiltext{3}{Harvard-Smithsonian Center for Astrophysics, 60 Garden Street, Cambridge, MA 02138, USA}
\altaffiltext{4}{Department of Astronomy, University of Washington, Box 351580, Seattle, WA 98195, USA}
\altaffiltext{5}{Max-Planck-Institut f\"ur extraterrestrische Physik, Giessenbachstrasse 1, D-85741 Garching, Germany} 
\altaffiltext{6}{Department of Astronomy and Astrophysics, University of California, 1156 High Street, Santa Cruz, CA 95064, USA}
\altaffiltext{7}{NSF Astronomy \& Astrophysics Postdoctoral Fellow.}
\altaffiltext{8}{Leopoldina Fellow.}

\begin{abstract}
We report on follow-up observations of candidate X-ray bright, radio-quiet isolated neutron stars (INSs) identified from correlations of the $\ROSAT$ All-Sky Survey (RASS) and the Sloan Digital Sky Survey (SDSS) Data Release 4 in \citet{me06}. We obtained {\it Chandra X-ray Telescope} exposures for $13$ candidates in order to pinpoint the source of X-ray emission in optically blank RASS error circles. These observations eliminated $12$ targets as good INS candidates. We discuss subsequent observations of the remaining candidate with the {\it XMM-Newton X-ray Observatory}, the Gemini North Observatory, and the Apache Point Observatory. We identify this object as a likely extragalactic source with an unusually high log $(f_X/f_{opt}) \sim 2.4$. We also use an updated version of the population synthesis models of \citet{Popov2009} to estimate the number of RASS-detected INSs in the SDSS Data Release 7 footprint. We find that these models predict $\sim$$3 -4$ INSs in the $11,000$ deg$^2$ imaged by SDSS, which is consistent with the number of known INSs that fall within the survey footprint. In addition, our analysis of the four new INS candidates identified by \citet{turner10} in the SDSS footprint implies that they are unlikely to be confirmed as INSs; together, these results suggest that new INSs are not likely to be found from further correlations of the RASS and SDSS.
\end{abstract}


\section{Introduction}
Nearly 15 years after the discovery with the {\it R\"ontgensatellit} ($\ROSAT$) by \citet{walter96} and \citet{haberl97} of RX J1856.4$-$3754 and RX J0720.4$-$3125, the first radio-quiet, X-ray bright isolated neutron stars (INSs), INSs remain rare. Only seven have been confirmed \citep[for reviews of the Magnificent Seven, see][]{haberl07,kaplan08}, confounding predictions that hundreds would be known by now \citep[e.g.,][]{treves91,blaes93}. Given the potential usefulness of X-ray observations of INSs in constraining the equation of state of matter at extreme densities, and the difficulties in constraining the Galactic population of INSs based on the current sample, there continue to be catalog-level attempts to identify new INSs. While several intriguing INS candidates have been identified recently, their exact nature remains open to interpretation \citep[e.g., Calvera;][]{rutledge08,hessels07,zane10} or their faintness at X-ray and optical wavelengths severely complicates their confirmation \citep[e.g., 2XMM J104608.7$-$594306;][]{pires09b,pires09a}. 

The Magnificent Seven were detected in $\ROSAT$ All-Sky Survey \citep[RASS;][]{voges99} data. Merging the RASS Bright and Faint Source Catalogs \citep[BSC, FSC;][]{voges99,fsc} yields $>$124,000 sources typically as faint as a few times $10^{-13}$~erg~cm$^{-2}$~s$^{-1}$. Identifying the counterparts to these sources (particularly those in the FSC) is on-going work, for which the Sloan Digital Sky Survey \citep[SDSS;][]{york00}, because of its large footprint, photometric depth, and spectroscopic follow-up, is an excellent tool. Correlations of the RASS and SDSS have produced studies of large numbers of common X-ray emitters, including main-sequence stars \citep{me09}, galaxies \citep{parejko08}, clusters \citep{popesso04}, and active galactic nuclei \citep[AGN;][]{anderson06}. The highest X-ray-to-optical flux ratios among these common X-ray emitters are typically measured for BL Lacs, for which log $(f_X/f_{opt}) < 2$ \citep[e.g.,][]{anderson06}. By contrast, the Magnificent Seven have log $(f_X/f_{opt})\sim4$ \citep{kaplan08}.

In \citet[][hereafter Paper I]{me06}, we used the merged BSC and FSC and an early version of the SDSS Data Release 4 \citep[DR4;][]{DR4paper} to identify candidate INS fields. Given the high $f_X/f_{opt}$ ratios expected for INSs, an optical counterpart to a new INS is likely to be well beyond the SDSS faint limit, $\sim$$22$~mag; none of the Magnificent Seven is brighter than $B = 25.2$ mag \citep{haberl07, schwope09}. We relied on SDSS and other archival data \citep[e.g., FIRST;][]{first} to identify plausible counterparts to RASS sources from among the categories described above and thereby reduced the number of X-ray error circles in which to search for new INSs. We excluded $99.9\%$ of the error circles in our sample and characterized the few surviving RASS fields as optically blank to the SDSS limit, implying that on average their log $(f_X/f_{opt}) > 1.6$. This was not intended to produce a complete sample of INS candidates, but rather to identify the candidates most worthy of follow-up. 

In Section~\ref{targets} we describe the INS candidates identified in Paper I observed with the {\it Chandra X-ray Observatory}; we discuss these observations briefly in Section~\ref{chandra_obs}. In Section~\ref{1406} we discuss observations of our best candidate after completion of the {\it Chandra} program, 1RXS J140654.5$+$525316, with the {\it XMM-Newton X-ray Observatory}, the Gemini Multi-Object Spectrograph (GMOS) on the Gemini North $8.1$-m telescope, Mauna Kea, HI, and the Seaver Prototype Imaging camera (SPIcam) on the Astrophysical Research Consortium $3.5$-m telescope at Apache Point Observatory, Sunspot, NM. In Section~\ref{pop} we use an updated version of the \citet{Popov2009} population synthesis model to estimate the number of expected INSs in the SDSS Data Release 7 \citep[DR7;][]{DR7paper} footprint; we also discuss the four new candidate INSs recently identified within the SDSS footprint by \citet{turner10}. We conclude in Section~\ref{concl}.

\begin{deluxetable*}{lclcc}
\tablewidth{0pt}
\tabletypesize{\scriptsize}
\tablecaption{Candidate INSs Observed With {\it Chandra}\label{chandra_targets}}
\tablehead{
\colhead{Source name} & \colhead{Minimum} & \colhead{Counterpart} & \colhead{Offset} & \colhead{X-ray} \\
\colhead{(1RXS J)}    &  \colhead{log $(f_X/f_{opt})$\tablenotemark{a}} & \colhead{(SDSS J)\tablenotemark{b}} &  \colhead{from RASS} & \colhead{Source ID}   
}
\startdata
003413.7$-$010134 & $1.8$ & 003413.04$-$010026.9 & 67\asec & QSO\\
092310.1$+$275448 & $1.5$ & 092314.20$+$275428.3 & 58\asec & QSO\\
102659.6$+$364039\tablenotemark{c} & $1.3$ & 102700.55$+$364016.0 & 26\asec & QSO\\
103415.1$+$435402 & $1.3$ & \nodata & \nodata & transient?\\
110219.6$+$022836 & $1.3$ & \nodata & \nodata & transient?\\
122344.6$+$373015 & $1.6$ & 122344.96$+$373019.3 & 8\asec & QSO?\\
140654.5$+$525316\tablenotemark{c} & $1.5$ & No SDSS counterpart & 9\asec & INS? \\
141944.5$+$113222\tablenotemark{c} & $2.6$ & \nodata & \nodata & transient? \\
142423.3$-$020201\tablenotemark{c} & $1.6$ & \nodata & \nodata & transient? \\
151855.1$+$355543 & $1.3$ & \nodata & \nodata & transient?\\
155705.0$+$383509\tablenotemark{c} & $1.3$ & \nodata & \nodata & transient? \\
162526.9$+$455750\tablenotemark{c} & $1.4$ & \nodata & \nodata & transient?\\
205334.0$-$063617\tablenotemark{c} & $1.9$ & \nodata & \nodata & transient?
\enddata
\tablenotetext{a}{Log $(f_X/f_{opt})$ is calculated as in \citet{macca88}. The listed value is for the brightest SDSS object in the error circle of radius $4\times$ the positional error of the $\ROSAT$ source, which was the area searched to identify plausible X-ray source counterparts. It is therefore a minimum value for the {\it true} RASS counterpart. See Paper I for details.}
\tablenotetext{b}{These are {\it Chandra}-detected sources with signal-to-noise ratios $>$3 that are plausible re-detections of the RASS sources; their SDSS counterparts' names are listed.}
\tablenotetext{c}{These sources were added to the original list of best candidates presented in Paper I; see text for details.}
\tablecomments{All targets were observed using the ACIS S3 chip and no grating. } 
\end{deluxetable*}

\section{INS candidates}\label{targets}
Our initial list of $11$ optically blank RASS fields is presented in Paper I. Among these were the field hosting the only confirmed INS in the DR4 footprint, RX J1605.3$+$3249, and the field of 1RXS J130547.2$+$641252, an INS candidate identified by \citet[][]{rutledge03} but rejected on the basis of their follow-up {\it Chandra} observations. After re-examining our list of INS candidates, three other sources were also eliminated from further consideration:
\begin{itemize}
\item 1RXS J013630.4$+$004226 and J131400.1$+$072312. Unidentified, X-ray emitting, faint optical clusters are the most likely ``ordinary'' RASS counterparts to survive our winnowing and contaminate our list of optically blank fields. Such clusters are difficult to identify from the SDSS data, and furthermore, the SDSS cluster catalog available to us\footnote{J.\ Annis, personal communication.} did not fully cover the DR4 area (1RXS J013630.4$+$004226 falls outside of this catalog's footprint). To estimate the likelihood that a candidate field hosts such a cluster, we measured the surface density of SDSS objects therein and compared the result to the distribution of surface densities for the \citet{popesso04} catalog of RASS/SDSS clusters. In these two fields, this comparison suggested that the RASS counterpart might be a faint cluster. Finally, the RASS images for 1RXS J013630.4$+$004226 reveal that it is detected in the hard but not the soft band, while J131400.1$+$072312 has a very low detection probability. Both are therefore poor candidates for follow-up observations.
\item 1RXS J141428.5$+$601707. A spectroscopically confirmed quasar (QSO) with an AGN-like $f_X/f_{opt}$ is on the edge of the error circle we searched for counterparts to the RASS source. Since the RASS images indicate that the source has no soft emission, this QSO cannot be ruled out as the RASS source counterpart. 
\end{itemize}

To the six remaining candidates we added seven intriguing fields. There were $13$ fields identified in Paper I as barely failing to meet our selection criteria for optically blank fields or as potentially hosting a faint optical cluster (see Table 2, Paper I). We reexamined these fields and determined that the following warranted new X-ray observations:

\begin{itemize}
\item 1RXS 162526.9$+$455750 and J205334.0$-$063617. These two fields met all of our selection criteria, but their cataloged positional error of $6\asec$ is likely underestimated, and they were therefore not included in the list of our best INS candidates.
\item 1RXS J140654.5$+$525316 and J142423.3$-$020201. These fields were not considered among our best INS candidates because of faint photometric candidate AGN at large angular separations from the RASS positions. However, as noted in Paper I, these SDSS objects appear unlikely to be the RASS sources. In the first case, the candidate AGN is $40\asec$ from the RASS position and has unreliable photometry, since it is fainter than the survey's 95\% completeness limit in its five ($ugriz$) bands. In the second case, multiple spectra obtained with the APO Double-Imaging Spectrograph\footnote{See \citet{me09} for a description of the observational set-up and spectral typing.} revealed that the candidate AGN is likely an ordinary, faint G star (its proper motion of $1.4$ mas yr$^{-1}$, measured by comparison to its USNO-B position, is further evidence of its stellar nature). 
\item 1RXS J141944.5$+$113222. This source was originally eliminated because the RASS images do not rule out that it is the same source as 1RXS J141949.0$+$113619; it otherwise met all our selection criteria.
\item 1RXS J102659.6$+$364039 and J155705.0$+$383509. These two sources were first identified as potential counterparts to faint optical clusters. Our later estimate of the surface densities of SDSS objects (described above), however, suggested that these two fields were not likely to host such clusters. Furthermore, 1RXS J102659.6$+$364039 is well detected in the RASS soft image; as for J155705.0$+$383509, its initial evaluation as a potential optical cluster counterpart had a relatively low confidence.\footnote{That is, the projected number of red sequence galaxies brighter than L$_{\star}$ within 1 Mpc is small (J.\ Annis, personal communication).}
\end{itemize}

\section{{\it Chandra} Observations}\label{chandra_obs}
The $13$ candidates were observed with the Advanced CCD Imaging Spectrometer \citep[ACIS;][]{burke1997} on board {\it Chandra} \citep{weisskopf1996}. We chose the S3 chip to image the sources because of its better low-energy sensitivity. The standard TIMED readout with a frame time of $3.2$ s was used, and the data were collected in VFAINT mode. In $12$ cases our {\it Chandra} observations led us to conclude that the RASS detection was not of a candidate INS (see Table~\ref{chandra_targets}; the Appendix includes a case-by-case discussion of these sources). 

\begin{figure*}[th]
\epsscale{1.15}
\centerline{\plotone{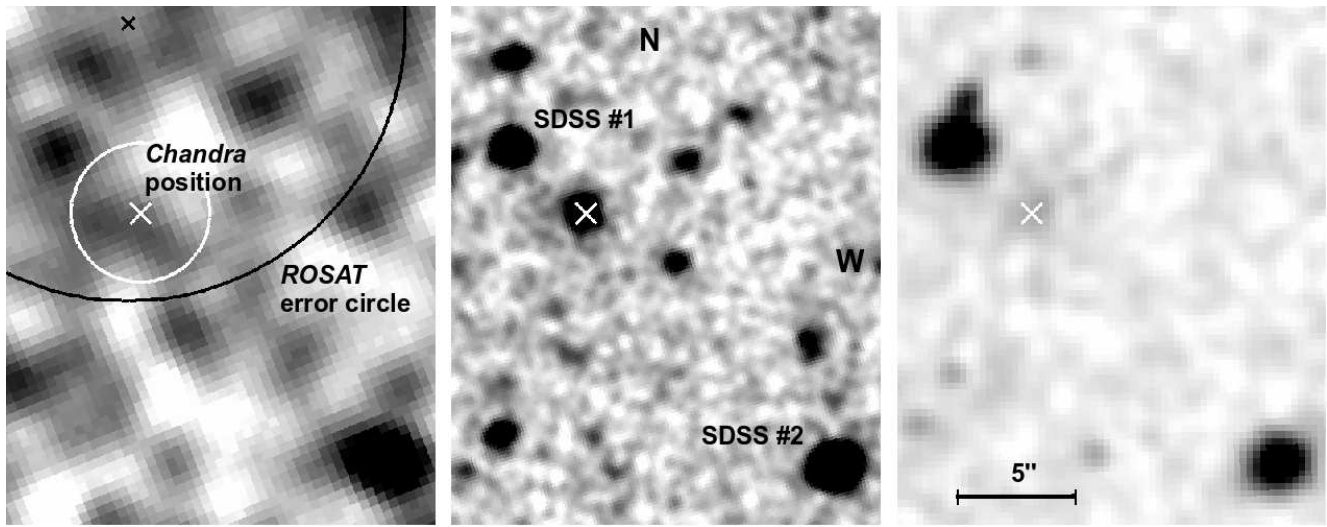}}
\caption{{\it Left} -- {\it Chandra} source position and $3\asec$ error circle overlaid on the SDSS $g$-band image for the field of 1RXS J140654.5$+$525316. The black cross and circle indicate the RASS source position and $12\asec$ error circle. {\it Middle} -- Co-added GMOS $g$-band image of the same field. The nearest SDSS objects to 1RXS J1406 with measured photometry are indicated. The {\it Chandra} source is clearly coincident with an object detected in the image. {\it Right} -- Co-added SPIcam $i$ image of the field. All three images are smoothed using a Gaussian function; the kernel radius is 5 pixels for the SDSS image and 3 for the two others. The stretch is the same for all three images.}\label{field}
\end{figure*}

\section{1RXS J140654.5$+$525316}\label{1406}
1RXS J140654.5$+$525316 (hereafter 1RXS J1406) is the only source for which we have a {\it Chandra} detection (with a signal-to-noise ratio $\sim$4) lacking an SDSS counterpart within the RASS field, and thus a plausible INS candidate. The {\it Chandra} source is $<$9\asec\ from the RASS position and offset $2.2$\asec\ from the observational aimpoint, implying that the $95\%$ encircled energy radius (at 1.5 keV) is $\sim$3\asec\ \citep{champ1}. The CIAO (version 4.0) task {\it celldetect} returns $23.9\pm6.3$ counts at the source position, and the {\it Chandra} and RASS data are therefore in good agreement: using {\tt WebPIMMS}, we find in both cases that the source $f_X$ ($0.1-2.4$~keV) $ \approx\ 1\times10^{-13}$~\ergcms, assuming a blackbody spectrum with a temperature of $90$~eV \citep[the median INS temperature;][]{haberl07}. 

In overlaying the {\it Chandra} and SDSS images, which are tied to the identical ICRS frame and have excellent astrometry (arcseconds and a few tenths of an arcsecond, respectively), we find that this source lacks a detectable optical counterpart in any of the SDSS $ugriz$ images within the {\it Chandra} error circle (see left panel, Figure~\ref{field}). A stack of the $gri$ SDSS images (the survey's deepest), which yields an effective imaging depth $\sim$23 mag ($3\sigma$), also shows no SDSS optical counterpart positionally consistent with the {\it Chandra} source. 

The closest optical object cataloged in SDSS, SDSS J140654.82$+$525310.7 (``SDSS \#1'' in Figure~\ref{field}), is offset by $4.5\asec$ from the {\it Chandra} position. This object is fainter in $ugr$ than the survey's $95\%$ completeness limits for point sources \citep{stoughton02}; its $i=21.06\pm0.10$ and $z=20.13\pm0.15$ mag are just below the limits in these bands.\footnote{Throughout, we use SDSS PSF magnitudes; PSF fitting provides better estimates of isolated star magnitudes. See \citet{stoughton02}.} At these magnitudes, the automated SDSS star/galaxy separation becomes unreliable \citep{scranton02}; indeed, the object is classified as both a star (in $ri$) and as a galaxy ($ugz$). 

We consider first the possibility that this object is a star and compare its $(r-i)$ and $(i-z)$ colors to those of main-sequence stars described in \citet{kev07}. Its colors are roughly consistent with those of an M4 star; we use $(z-J) = 1.56$ for these stars \citep{kev07} to estimate that the star's $J\approx18.5$ mag. We then calculate log $(f_X/f_J)$ as in \citet{me09}. This returns a (log) flux ratio of $\sim$$0.8$, which is (unsurprisingly) much higher than is usually seen for SDSS M star counterparts to $\ROSAT$ sources, which have log $(f_X/f_J) < 0$ \citep[see Figure 5 in][]{me09}. The large optically inferred distance for such a star (several kpc) further argues against it being the $\ROSAT$-detected source, as does the corresponding $L_X > 10^{31}$ \ergcms\ \citep[to compare this to ``ordinary" RASS-detected stars, see, e.g.,][]{me09}.

If SDSS J140654.82$+$525310.7 is instead a background galaxy, its $(u-g) = -0.85\pm0.55$ is well below the typical $(u-g) = 0.6$ threshold commonly used to identify low-redshift SDSS QSOs. However, such a blue $(u-g)$ color is beyond what is typically seen for even the bluest SDSS objects, and the galaxy's position elsewhere in color space is inconsistent with it hosting a QSO \citep{richards02}. A more likely explanation for this anomalously large UV-excess is the red leak of the SDSS $u$-band filter, which is worst for the reddest objects.\footnote{See {\tt http://www.sdss.org/dr7/products/catalogs/index.html}.} Ordinary galaxies are weak X-ray emitters, and we conclude that SDSS J140654.82$+$525310.7, whether star or galaxy, is unlikely to be the {\it Chandra}/RASS counterpart.

The {\it Chandra} observations therefore confirm 1RXS J1406 as a good candidate INS, since we do not expect any such object to have a cataloged counterpart in SDSS. Furthermore, based on the absence of an SDSS counterpart, the optical counterpart to 1RXS J1406 has log $(f_X/f_{opt})\ \gapprox\ 2$. Among ``ordinary'' objects known to have high flux ratios, virtually none are $>$100$\times$ brighter in the X ray than in the optical \citep[][]{stocke91, zickgraf03, anderson06}, motivating our follow-up observations of 1RXS J1406.

\subsection{{\it XMM-Newton} Observations}
To test whether 1RXS J1406 has a blackbody spectrum peaking in soft X rays ($40 - 100$ eV), as is the case with known INSs \citep{haberl07}, we observed the source with {\it XMM} on 2007 Jun 21 (ObsId $0503960101$). Data were obtained from the two MOS CCDs \citep{turner01} and the pn CCD \citep{struder01}. The CCD pixel sizes are $1.1$ and $4.1\asec$, respectively, while the mirror point spread function is $\sim$6\asec\ full-width half maximum (FWHM). We observed with the thin filter and in full-frame mode. The time resolution of EPIC-pn in this mode, 70\,ms, was expected to be sufficient to see prominent pulsations, assuming 1RXS J1406 resembles the six INSs for which periods are known \citep[these have periods between $3-11$ s;][]{kaplan09}.

\begin{deluxetable*}{lcccc}[!t]
\tablewidth{0pt}
\tabletypesize{\scriptsize}
\tablecaption{Model fits to the {\it XMM} spectrum of 1RXS J140654.5$+$525316 \label{xmmtable}}
\tablehead{
\colhead{} & \colhead{Reduced}  &\colhead{$N_H$} & \colhead{} & \colhead{Flux, $0.1 - 4$ keV} \\
\colhead{Model} & \colhead{${\chi^2}$} & \colhead{(cm$^{-2}$)}& \colhead{$\Gamma$ or $kT$} & \colhead{(\ergcms)}
}
\startdata
power law      & $0.84$ & $1.9^{+0.8}_{-0.6} \times 10^{21}$ & $1.94^{+0.3}_{-0.2}$ & $9.4 \times 10^{-14}$\\
bremsstrahlung & $0.89$ & $1.3^{+0.5}_{-0.4} \times 10^{21}$ & $3.42^{+2.4}_{-1.1}$ & $9.3 \times 10^{-14}$
\enddata
\tablecomments{For both models the number of degrees of freedom is 36. In the fourth column we give the photon index $\Gamma$ of the power law and temperature in keV of the bremsstrahlung model. The quoted errors correspond to the 90\% confidence levels. Fluxes include absorption by the interstellar medium.}
\end{deluxetable*}

The data were reduced with the Standard Analysis Software (SAS\footnote{{\tt http://xmm.vilspa.esa.es}.} version 8.0.0). Roughly $30\%$ of the $\sim$25 ks observation was lost due to background flaring. Good time intervals were selected from binned background light curves. 100~s bins with more than 35 and 40 counts were rejected for the MOS1/2 and pn data respectively, reducing the effective exposure times to 18 and 12~ks. We applied the SAS task {\it epreject} to the pn data to correct effects in the offset map caused by particle events. For the spectral analysis we used only single and double events in the pn data, and single, double, triple, and quadruple events in the MOS1/2 data. We filtered the data to exclude bad pixels and CCD gaps.

The spectral analysis was restricted to events with energies $0.3 - 4$ keV for the MOS1/2 data and $0.13 - 4$ keV for the pn data. We find that the count rates for 1RXS J1406 are $1.1\pm0.1 \times$\ctss\ (MOS1/2) and $3.9\pm0.2 \times$\ctss\ (pn); given the errors and slightly different energy ranges, this is consistent with the source being non-variable when comparing the {\it XMM} and $\ROSAT$/{\it Chandra} count rates. 

The source spectra were extracted from the event files in a region centered on the source and extending $30\asec$ in radius and binned to have $\geq25$ counts per bin. (Spectra for a nearby source-free region of the same size were extracted to estimate the background.) Model spectra were fitted simultaneously to data from the three {\it XMM} detectors using XSPEC (version 12) with Churazov weighting. We used the \citet{wilms2000} {\it tbabs} model for interstellar absorption and tested the following models individually and in combinations: power law, bremsstrahlung, blackbody, and neutron star atmosphere (NSA).

\begin{figure}[th]
\epsscale{1.1}
\centerline{\includegraphics[width=1\columnwidth]{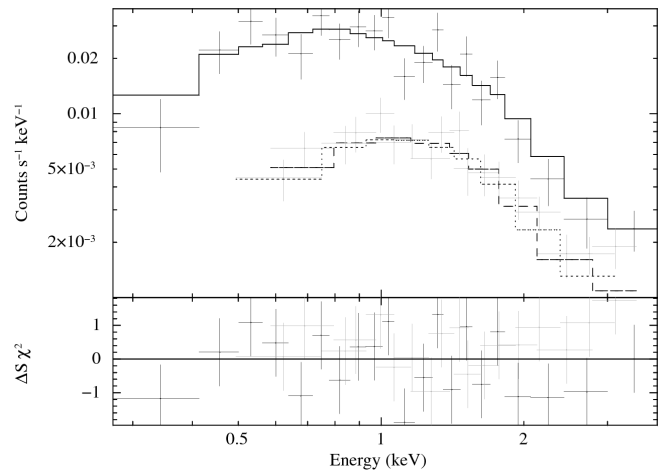}}
\caption{Power-law model fit to the pn (black crosses and solid line) and MOS1 and 2 (light gray crosses, dashed and dotted lines, respectively) data for 1RXS J140654.5$+$525316. (A color version of this figure is available in the online journal.)}\label{xmmfit}
\end{figure}

The power-law and bremsstrahlung models represent the observed spectrum comparably well ($\chi^2= 30.14$ for 36 degrees of freedom for the power-law model, $\chi^2= 31.91$ for 36 degrees of freedom for the bremsstrahlung), while a simple blackbody did not fit the data well ($\chi^2= 61.18$ for 36 degrees of freedom). A composite power-law and blackbody model resulted in only an upper limit for the blackbody contribution, indicating that no second component is required in addition to the power law. 

To test the NSA models, we set the NS mass and radius to the canonical values, 1.4 $M_{\odot}$ and 12 km, and varied the magnetic field strength. This produced fits with unreasonably large estimates for the distance to the source and unreasonably small estimates for the line-of-sight $N_H$, both with very large errors. (Freezing the column density to the Galactic value resulted in significantly worse fits.)

The fit parameters for the two best single-component models are listed in Table~\ref{xmmtable}; the data and best-fit power-law spectrum and residuals are shown in Fig.~\ref{xmmfit}. The $N_H$ derived from this fit is $\sim$10$^{21}$\,cm$^{-2}$; the Galactic value as obtained from HI surveys \citep{Kalberla2005,Dickey1990} in this direction is only $\sim$10$^{20}$\,cm$^{-2}$, suggesting that there is some local absorption at the source.
 
Finally, the $0.15 - 2$ keV pn light curves were checked for periodic variations using the FTOOLS\footnote{{\tt http://heasarc.gsfc.nasa.gov/docs/software/ftools/
ftools\_menu.html}.} {\it powspec} and {\it efsearch}. No significant signal was found in the period range from 200 ms to 10 ks: assuming a sinusoidal variation we obtained a 3$\sigma$ upper limit of $30\%$ for the pulsed fraction.

Together, these pieces of evidence -- the hard X-ray spectrum, the spectral shape, the high derived $N_H$ value, the absence of pulsations -- especially when combined with its very faint optical counterpart (see below), suggest
that 1RXS J1406 is an AGN.

\subsection{Deeper Optical Imaging}
Given the $\sim$22 mag limit of the SDSS survey and the faintness of known optical counterparts to INSs, we obtained deeper optical data of the 1RXS J1406 field.

\subsubsection{Gemini}
We used the Gemini Multi-Object Spectrograph \citep{hook2004} on the Gemini North telescope to obtain optical images of the 1RXS J1406 field (proposal GN-2008A-Q-107-1). GMOS consists of a row of three $2048 \times 4608$ pixel CCDs with $\sim$0.5 mm gaps between them; the field of view is $5.5\times5.5$ arcmin$^2$. Five Gemini queue mode observations were carried out on 2008 Jan 17, May 3 (three), and May 4; each observation was $750$ s long. The May nights were dark and photometric, with a median seeing of 0.6\asec\ or slightly better (measured from the FWHM of $\sim$15 stars in the individual images); the Jan night was gray and not photometric, with seeing $\sim$ 0.7\asec. The images were taken with a modified SDSS $g$ filter centered at $4750$ \AA\ that provides coverage over the $3360-3850$ \AA\ range. We used $2\times2$ binning, so that the plate scale is $0.1454\asec$ per pixel. 

The raw images were processed using the Gemini external package for IRAF\footnote{IRAF is distributed by the National Optical Astronomy Observatories, which are operated by the Association of Universities for Research in Astronomy, Inc., under cooperative agreement with the National Science Foundation.} (version 1.9). An overscan-subtracted and trimmed master bias was recreated with {\it gbias} from the 64 biases used to create the master bias distributed with the data, which is not overscan-subtracted. Separate overscan- and bias-subtracted trimmed master flats were created with {\it giflat} for the two observational epochs. The master flat applied to the Jan 17 observation was produced from $16$ flats taken 2008 Jan 14, and the one applied to the May observations from $13$ flats taken 2008 Apr 22, in the same instrumental configuration as our observations.

The individual observations were overscan- and bias-subtracted, trimmed, and flat-field corrected using {\it gireduce} before being mosaiced with {\it gmosaic}. The mosaiced images were inspected before being co-added using {\it imcoadd}. The Jan 17 image has a background level that is several times that of the May images, and given the (relatively) inferior image quality for this observation,\footnote{For definitions of Gemini image quality, see {\tt http://www.gemini.edu/sciops/telescopes-and-sites/observing-
condition-constraints}.} we did not include it in creating the final co-added image used for all the subsequent analysis. 

We used the SDSS DR7 photometric catalog and the \textsl{Graphical Astronomy and Image Analysis Tool} \citep[$GAIA$;][]{gaia} to correct the astrometry of this co-added image. For bright stars, the DR7 astrometry has statistical errors per coordinate of $\sim$45 mas, with systematic errors of $<$20 mas \citep{DR7paper}. The applied correction was $\sim$1.6\arcsec; with this correction, 1RXS J1406 is clearly positionally coincident with an object in the co-added GMOS image (see middle panel, Figure~\ref{field}).

To determine the magnitude of this object, we first used the IDL routine {\it FIND} to identify objects in the co-added image. Separately, we queried the SDSS Skyserver\footnote{\tt http://cas.sdss.org/astrodr7/en/tools/search/IQS.asp.} for objects within 5\amin\ of the position of 1RXS J1406 with $13 < g < 23$ mag. These two lists were matched, and photometry was extracted at the positions of the matched objects in the co-added image using the IDL routine {\it APER} and an aperture radius of 12 pixels, which is slightly more than twice the estimated FWHM for objects in the image. The 71 objects for which {\it APER} returned a magnitude had a median $g$ offset of $9.78\pm0.13$ mag relative to their SDSS PSF magnitudes. 

We then use {\it APER} to extract a magnitude for the GMOS-detected counterpart to 1RXS J1406. Applying the photometric offset determined above, we find that its $g = 24.75\pm0.14$ mag. This implies that 1RXS J1406 has log $(f_X/f_{opt}) \sim 2.4$. Although this flux ratio is unusually high for most typical X-ray-emitting subclasses, and the 1RXS J1406 counterpart may well have an interesting nature, this ratio is too low for a plausible INS identification.

\subsubsection{Apache Point Observatory}
We also obtained $i$-band images (centered at $7700$ \AA) of the 1RXS J1406 field on 2008 May 2 with the Seaver Prototype Imaging camera on the 3.5-m telescope at Apache Point Observatory. The SPIcam detector is a backside-illuminated $2048 \times 2048$ pixel CCD that produces images with a plate scale of $0.28\asec$ per pixel and a field of view of $4.8\times 4.8$ arcmin$^2$. The dithered images were reduced, aligned, and combined using standard IRAF routines: {\it zerocombine}, {\it flatcombine}, {\it imalign}, and {\it imcombine}. The exposures were either 180 or 300 s; we co-added the 12 best images, which were taken with seeing better than $1.7$\asec.

The astrometry of our co-added SPIcam image was corrected as described above; the {\it GAIA}-processed image was shifted by $\sim$17\arcsec\ relative to the original (see right panel, Figure~\ref{field}). We once again matched the output from {\it FIND} to the SDSS reference stars and then extracted the photometry for the resulting 39 objects from the co-added image using {\it APER} (the aperture radius here was 10 pixels). Relative to their SDSS $i$ magnitudes, the median offset for these objects was $5.57\pm0.09$ mag. We then find that the counterpart to 1RXS J1406 has $i = 22.59\pm0.16$ mag and $(g-i) = 2.16\pm0.21$.

Combined, the {\it XMM} and new optical data suggest that 1RXS J1406 is a likely optically faint, X-ray bright AGN whose X-ray spectrum is modeled by a power law with a photon index $\Gamma \sim 2$ and whose $(g-i)$ color is consistent with a $z \sim 4$ QSO \citep{richards02}. This implies an X-ray luminosity $>$$10^{46}$ \ergs, at the high end for what is seen for e.g., broad-line AGN in the {\it Chandra} Multiwavelength Project \citep{silverman05}. 

\section{How many INSs are there in the SDSS DR7 footprint?}\label{pop}
\subsection{Population synthesis estimates}
The dearth of new, confirmed INSs despite a number of systematic searches \citep[in addition to our own, e.g.,][]{rutledge03} has motivated theoretical work to reconcile predictions and observations. Models of interstellar accretion onto INSs, once thought to be the main mechanism for reheating older NSs, thereby making them detectable, now suggest that most such objects will in fact be invisible to current X-ray telescopes \citep{perna03, ikhsanov}. Separately, improving Galactic population models for young, cooling INSs indicate that at the bright end of the X-ray log N-log S distribution, where most INS searches have taken place, the current number of known INSs is consistent with expectations.

We used the recent population synthesis calculations of \citet{Popov2009} to estimate the number of cooling INSs expected to fall within the SDSS DR7 footprint. The DR7 imaging area is over $11,000$ deg$^2$, or nearly twice that of DR4; while the bulk of that additional sky coverage is around the North Galactic Cap, the DR7 footprint also includes a number of stripes, imaged as part of the Sloan Extension for Galactic Understanding and Exploration (SEGUE) survey \citep{SEGUEpaper}, which pass through the Galactic Plane. Of the magnetic field distributions discussed in \citet{Popov2009}, we use the lognormal distribution labeled G3. The parameters of this magnetic field distribution are closest to the ones of the model found to be consistent simultaneously with the observed populations of cooling neutron stars, radio pulsars, and magnetars. 

The population synthesis code produces Galactic maps that predict the number of INSs for specific count rate intervals \citep[for details, see][]{bettina08}. Because our search for new INS candidates merged the RASS BSC and FSC to create a source catalog, the depth of our X-ray ``survey'' varies across the sky.\footnote{See the RASS exposure map at {\tt http://www.xray.mpe.mpg.de/rosat/survey/rass-3/main/help.html}.} To approximate the count rate limit reached at each position we assume a minimum of 6 counts, which is the threshold for inclusion in the FSC \citep{fsc}. However, in the FSC this cut is imposed after X-ray background subtraction, which we do not include; as a result, we slightly overestimate the number of expected RASS INSs.

We used the stripe definitions available from the SDSS website to construct a version of the DR7 footprint that is uniquely defined, in the sense that areas where the SDSS and SEGUE stripes overlap are included only once. These stripes were then tested to ensure that SDSS photometric data were available for points at 1 deg intervals along the stripes, as a number of stripes have gaps.\footnote{See {\tt http://www.sdss.org/DR7/coverage/}.} The resulting footprint was transformed into a mask in Galactic coordinates for use with the population synthesis code; due to computational issues (finite grid size, re-sampling) the footprint in Galactic coordinates is slightly less than $11,000$ deg$^2$.

By combining the DR7 footprint mask with the RASS count-rate-minimum map, we obtained a histogram representing the range in X-ray sensitivity across the DR7 area. The lowest count rate limit (and thus the deepest observation) is $1.7 \times 10^{-4}$ count s$^{-1}$, while the highest count rate limit, due to a very short exposure time, is 6383 count s$^{-1}$. To sample this wide distribution of count rates, we chose bins of varying size when running the population synthesis code, so that we produced 12 maps for count rates between $1.7 \times 10^{-4}$ and $3 \times 10^{-2}$ count s$^{-1}$, one map for count rates between $3 \times 10^{-2}$ and $5 \times 10^{-2}$ count s$^{-1}$ (which is the faint limit of the BSC), and one map for $>$5$\times 10^{-2}$ count s$^{-1}$.

The population synthesis maps appropriate for the count rate limit at a given position in the footprint are added up to derive the predicted number of RASS-observable INSs for those coordinates. Finally, the sum over the entire footprint area is computed. Depending on which model is chosen when estimating absorption due to the interstellar medium \citep[ISM; for details about these models, see][]{bettina08}, we predict that 3.7 (old analytical ISM model) to 3.9 (Hakkila-ISM model) INSs detectable by $\ROSAT$ fall in the DR7 area. This prediction changes very little if we consider only count rates down to $2\ \times$~\ctss. As the effect of ignoring the X-ray background is most likely to be largest when estimating the lowest possible count rates, we expect that our predicted number of INSs is only a slight overestimate. 

This prediction is consistent with current observations: of the Magnificent Seven, three are within the DR7 footprint: RX J1308.6$+$2127, RX J1605.3$+$32491 (the only INS in the DR4 footprint), and RXS J214303.7$+$065419. This suggests that uncovering new INSs from correlations of the RASS and SDSS is, at best, unlikely.

\subsection{The \citet{turner10} candidates}\label{comp}
\citet{turner10} recently identified a number of new candidate INSs in the sky above $-39^{\circ}$ based on a search of the RASS BSC. These are X-ray sources for which these authors first found no statistically plausible counterpart in the optical, infrared, or radio, through correlations with the USNO-A2, IRAS, and NVSS catalogs. Of the roughly 150 sources with a probability $\geq$80\% of being unassociated with an optical/infrared/radio source, \citet{turner10} then obtained {\it Swift}/XRT observations for close to 100 in order to decrease their positional uncertainties to $\sim$3.5\asec\ before repeating the search for counterparts. The nine surviving INS candidates have no counterparts in USNO-A2, 2MASS, NVSS, IRAS, or in the {\it Swift} UVOT observations made simultaneously with the X-ray observations (cf.\ their Table 5). Of these nine, four fall within the current SDSS footprint; these four also fall within the DR4 footprint searched in Paper I, and did not make our list of best candidate INSs. We discuss below why none of these sources met our original search criteria, as well as what information can be gleaned about their nature from correlations with SDSS of the subsequent \citet{turner10} {\it Swift} observations of the BSC sources.

\begin{itemize}
\item In identifying our candidate INSs, we eliminated all RASS sources for which we found a nearby SDSS object with a UV-excess. These are mostly candidate (photometric) QSOs \citep[e.g.,][]{richards02}, but this cut also removed white dwarfs, cataclysmic variables, and X-ray binaries. In practice, we removed from consideration any RASS source for which we found an SDSS object offset from the X-ray positions by less than 4$\times$ the quoted RASS positional uncertainty and satisfying $(u-g) < 0.6$ and $u \leq 22.0$ mag. The \citet{turner10} candidate INS 1RXS J130205.2$+$155122 is within 12.5\asec\ of an SDSS object with a UV-excess, SDSS J130205.19$+$155134.4 ($(u-g) = 0.52\pm0.13$, $u=21.60\pm0.13$). This offset corresponds to less than twice the quoted RASS positional uncertainty of 9\asec, and this source was therefore removed from our candidate list. Later SDSS spectroscopy confirmed that SDSS J130205.19$+$155134.4 is a $z=0.534\pm0.001$ QSO, and it is listed as the counterpart to 1RXS J130205.2$+$155122 in the \citet{anderson06} catalog of RASS/SDSS AGN. 

SDSS J130205.19$+$155134.4 is positionally coincident (within $0.5$\asec) with the {\it Swift} source detected by \cite{turner10}, XRT J130205.2$+$155134.0, and it is very probable that this QSO is the BSC/{\it Swift} source counterpart.

\item 1RXS J144359.5$+$443124 is less than 11\asec\ away from a UV-excess object, SDSS J144400.25$+$443117.9 ($(u-g) = 0.24\pm0.13$, $u=21.70\pm0.13$). This object lacks SDSS spectroscopy, but its proximity (the positional uncertainty for this source is 10\asec) and X-ray-to-optical flux ratio (if it is the RASS counterpart) of log $(f_X/f_{opt})$ = 1.7 suggest that SDSS J144400.25$+$443117.9 is plausibly a QSO and the RASS source counterpart. Furthermore, SDSS J144400.60$+$443119.5, another nearby object with a UV-excess $(u-g) = 0.85\pm0.29$ (but too faint to meet the cut described above, as its $u=22.62\pm0.29$), has an intriguing emission-line spectrum. While the SDSS spectrum is not definitive, this object is a candidate Seyfert-type galaxy, and therefore another possible counterpart to the RASS source (it is offset by less than 13\asec\ from the X-ray position) with log $(f_X/f_{opt}) = 1.8$. The presence of either of these objects within the RASS field would be sufficient for the latter to be removed from our list of candidate INSs.

SDSS J144400.60$+$443119.5 is offset by 4.8\asec\ from XRT J144400.5$+$443124.2, the {\it Swift} source detected by \cite{turner10}, making this SDSS object a plausible counterpart to the BSC/{\it Swift} source. There is also a very faint UV-excess object offset from the {\it Swift} position by 3.4\asec\ that is a possible source counterpart, but with $u=23.65\pm0.60$ and $g=23.00\pm0.13$ mag it is fainter than the SDSS 95\% completeness limit and therefore requires deeper imaging to obtain reliable colors; it is also fainter than the spectroscopic survey limit.

\item 1RXS J212700.3$+$101108 is 30\asec\ away from another UV-excess object, SDSS J212658.27$+$101105.2 ($(u-g) = 0.28\pm0.08$, $u=20.68\pm0.07$). This object also lacks SDSS spectroscopy, but here again, proximity (the positional uncertainty is 12\asec) and flux ratio (log $(f_X/f_{opt})$ = 1.3) are consistent with SDSS J212658.27$+$101105.2 being a RASS-detected QSO. 

The {\it Swift} source XRT J212700.3$+$101122.2 is within $3.5$\asec\ of a different UV-excess object, SDSS J212700.20$+$101119.1 ($(u-g) = 0.30\pm0.16$, $u=21.80\pm0.16$),with a log $(f_X/f_{opt})$ = 1.7; while this object lacks spectroscopy, it is a plausible QSO-counterpart to the X-ray source.

\item 1RXS J230334.0$+$152019 is within 18\asec\ of yet another UV-excess object, SDSS J230332.79$+$152015.3 ($(u-g) = 0.54\pm0.16$, $u=21.65\pm0.16$). This object also lacks SDSS spectroscopy, but its proximity to the RASS source (the positional uncertainty is 11\asec) and log $(f_X/f_{opt})$ = 1.6 are consistent with SDSS J230332.79$+$152015.3 being a RASS-detected QSO. We also note that the RASS flag for this source is set to $1$, indicating that the X-ray data are not reliable \citep[see][]{voges99}. Furthermore, visual inspection of the RASS image suggests that there are two sources in the field, as does the pointed $\ROSAT$ High Resolution Imager observation of the same field, although in neither case was the fainter source cataloged. This would have also eliminated this source from our consideration, as we inspected our candidates to remove possible artifacts and extended or very uncertain X-ray detections.

1RXS J230334.0$+$152019 has not been re-observed with {\it Swift}.
\end{itemize}

In summary, while spectroscopic follow-up of the UV-excess SDSS objects in three of these RASS fields is certainly needed to confirm their nature, re-applying the criteria used in Paper I to the four \citet{turner10} candidates for which SDSS data are available suggests that these authors' search is also unlikely to add to the number of INSs falling within the SDSS footprint. Updating the BSC positions to those of the {\it Swift} detections and matching these to the SDSS catalogs only increases the likelihood that the \cite{turner10} candidate INSs are in fact X-ray-detected QSOs.

\section{Conclusion}\label{concl}
We have used follow-up observations of candidate INSs identified from correlations of the RASS and SDSS to determine that none is a likely new INS. Our best candidate, 1RXS J140654.5$+$525316, is a likely extragalactic source with a high log $(f_X/f_{opt}) \sim 2.4$. (The very large sample of X-ray emitting AGN assembled by \citet{anderson06} includes only a handful of objects out of $\sim$7000 with a ratio this large, indicating that this object may be worthy of further study in its own right.) Applying an updated version of the population synthesis models of \citet{Popov2009} to estimate the number of RASS-detected INSs in the SDSS DR7 footprint, we find that these models predict $\sim$$3 - 4$ INSs in the $11,000$ deg$^2$ imaged by the survey. This is consistent with the number of known INSs that fall within the survey footprint. Furthermore, our analysis of the four INS candidates recently identified by \citet{turner10} for which there are SDSS data implies that none is likely to be confirmed as a new INS. These results suggest that new INSs are unlikely to be found from further correlations of the RASS and SDSS. This (probable) absence of new INSs in the SDSS footprint is unsurprising in light of the predictions of \citet{bettina08}: while SDSS focused on the North Galactic Cap, new INSs are more likely to be found in the Galactic plane.

These new INSs are expected to be young and hot, but also farther away than the seven $\ROSAT$ sources, rendering the confirmation of any candidate INSs very challenging \citep[see, e.g.,][]{pires09b,pires09a}. There is some hope that the eROSITA instrument aboard the Russian satellite {\it Spectrum-X-Gamma}, planned to launch in 2012, will be able to detect these objects because of its sensitivity at low energies \citep{erosita}. However, the instrument's point spread function ($15$\asec) and angular resolution ($28$\asec) suggest that identifying new INSs in its data will not be trivial. The proposed Wide-Field X-ray Telescope, dedicated to performing surveys of the sky in the soft X-ray band ($\sim$$0.4-6$~keV) and with the ability to resolve sources with a resolution 20$\times$ that of $\ROSAT$, may well uncover hundreds more INSs \citep{campana10}. Similar predictions, however, were made before the launch of $\ROSAT$ \citep[e.g.,][]{treves91,blaes93} and have yet to be realized. The Magnificent Seven do not seem likely to turn into a Dirty Dozen any time soon. 

\section*{Acknowledgments}
M.A.A.\ is supported by an NSF Astronomy and Astrophysics Postdoctoral Fellowship under award AST-0602099. B.P.\ acknowledges the support of the Deutsche Akademie der Naturforscher Leopoldina (Halle, Germany) under grant BMBF-LPD 9901/8-170. We gratefully acknowledge {\it Chandra} grant GO6-7058X and NASA award NNX07AO59A for support of portions of our program. We thank Tom Matheson, Dara Norman, and Anil Seth for their help with the Gemini data,  Michael Strauss and Chris Beaumont for their help reconstructing the SDSS footprint, and Paul Green for useful discussions about 1RXS J1406. We thank the referee for suggestions that improved the manuscript.

This research has made use of software provided by the {\it Chandra} X-ray Center in the application packages CIAO, ChIPS, and Sherpa. Observations were obtained with {\it XMM-Newton}, an ESA science mission with instruments and contributions directly funded by ESA Member States and NASA.

Funding for the SDSS and SDSS-II has been provided by the Alfred P. Sloan Foundation, the Participating Institutions, the National Science Foundation, the U.S. Department of Energy, the National Aeronautics and Space Administration, the Japanese Monbukagakusho, the Max Planck Society, and the Higher Education Funding Council for England. The SDSS Web Site is {\tt http://www.sdss.org/}.

The SDSS is managed by the Astrophysical Research Consortium for the Participating Institutions. The Participating Institutions are the American Museum of Natural History, Astrophysical Institute Potsdam, University of Basel, University of Cambridge, Case Western Reserve University, University of Chicago, Drexel University, Fermilab, the Institute for Advanced Study, the Japan Participation Group, Johns Hopkins University, the Joint Institute for Nuclear Astrophysics, the Kavli Institute for Particle Astrophysics and Cosmology, the Korean Scientist Group, the Chinese Academy of Sciences (LAMOST), Los Alamos National Laboratory, the Max-Planck-Institute for Astronomy (MPIA), the Max-Planck-Institute for Astrophysics (MPA), New Mexico State University, Ohio State University, University of Pittsburgh, University of Portsmouth, Princeton University, the United States Naval Observatory, and the University of Washington.

\renewcommand{\thesection}{A\arabic{section}}
\setcounter{section}{0}  
\section*{Appendix: Chandra Observations Of INS Candidates}
In $12$ of our candidate INS fields observed by {\it Chandra} we failed to re-detect the RASS source or identified it as an ordinary X-ray emitter. In using the RASS count rates to predict the exposure lengths required to detect these sources with {\it Chandra}, we assumed that the sources are $90$~eV blackbodies \citep[the median INS temperature;][]{haberl07}; the predicted {\it Chandra} rates were about 10\% higher than the RASS count rates (see Table~\ref{chandra_details}). Source lists for these fields were produced by the standard {\it Chandra} data pipeline using {\it celldetect} with a signal-to-noise threshold for detections set to 3; for comparison, we also generated {\it celldetect} source lists setting this threshold to 2.

\begin{deluxetable*}{lccclcc}
\tablewidth{0pt}
\tabletypesize{\scriptsize}
\tablecaption{{\it Chandra} Observation Details\label{chandra_details}}
\tablehead{
\multicolumn{4}{c}{{\it ROSAT}} & \multicolumn{3}{c}{{\it Chandra}}\\
\colhead{Source name} & \colhead{$1\sigma$} & \colhead{Detection} & \colhead{Count Rate} & \colhead{Observation} & \colhead{Exposure}   & \colhead{} \\
\colhead{1RXS J}      & \colhead{(\asec)}   & \colhead{Likelihood} & \colhead{(\ctss)}   & \colhead{Date}        &  \colhead{Time (ks)} & \colhead{ObsId}   
}
\startdata
003413.7$-$010134 & $14$ & $10$ & $1.3\pm0.6$ & 2006 May 14 & 2.0 & \dataset[ADS/Sa.CXO#Obs/6693]{6693}\\
092310.1$+$275448 & $14$ & $14$ & $2.5\pm1.0$ & 2006 Jan 2  & 1.0 & \dataset[ADS/Sa.CXO#Obs/6701]{6701}\\
102659.6$+$364039 & $12$ & $26$ & $6.1\pm1.6$ & 2006 Jun 13 & 1.0 & \dataset[ADS/Sa.CXO#Obs/6702]{6702}\\
103415.1$+$435402 & $14$ & $9$  & $1.7\pm0.8$ & 2006 Mar 13 & 1.5 & \dataset[ADS/Sa.CXO#Obs/6696]{6696}\\
110219.6$+$022836 & $15$ & $11$ & $1.8\pm0.8$ & 2006 Jun 14 & 1.5 & \dataset[ADS/Sa.CXO#Obs/6697]{6697}\\
122344.6$+$373015 & $15$ & $11$ & $2.8\pm1.1$ & 2006 Mar 13 & 1.0 & \dataset[ADS/Sa.CXO#Obs/6695]{6695}\\
140654.5$+$525316 & $12$ & $10$ & $1.3\pm0.6$ & 2006 Sep 8  & 2.0 & \dataset[ADS/Sa.CXO#Obs/6703]{6703}\\
141944.5$+$113222 & $6$  & $7$  & $2.2\pm1.1$ & 2006 Aug 21 & 1.5 & \dataset[ADS/Sa.CXO#Obs/6700]{6700}\\
142423.3$-$020201 & $6$  & $8$  & $2.6\pm1.1$ & 2006 Jun 13 & 1.0 & \dataset[ADS/Sa.CXO#Obs/6694]{6694}\\
151855.1$+$355543 & $9$  & $32$ & $3.3\pm0.9$ & 2006 Sep 16 & 1.0 & \dataset[ADS/Sa.CXO#Obs/6705]{6705}\\
155705.0$+$383509 & $15$ & $9$  & $3.0\pm1.4$ & 2006 Aug 12 & 1.0 & \dataset[ADS/Sa.CXO#Obs/6699]{6699}\\
162526.9$+$455750 & $6$  & $7$  & $1.4\pm0.6$ & 2006 Jul 31 & 2.0 & \dataset[ADS/Sa.CXO#Obs/6704]{6704}\\
205334.0$-$063617 & $6$  & $9$  & $2.1\pm0.8$ & 2005 Dec 14 & 1.5 & \dataset[ADS/Sa.CXO#Obs/6698]{6698}
\enddata
\end{deluxetable*}

\section*{Candidate RASS counterparts at large offsets}
In four fields we detect potential counterparts to the RASS source and identify its SDSS counterpart. In all four cases, lowering the detection threshold fails to uncover any {\it Chandra} sources closer to the RASS position. 

\begin{itemize}
\item 1RXS J003413.7$-$010134: A {\it Chandra} source ($1.7\pm0.4 \times$\ctss) is detected $67$\asec\ from the RASS position. This source is positionally coincident (separation $<1\asec$) with a spectroscopically confirmed $g=17.24\pm0.02$ mag QSO with $z = 1.292\pm0.002$, SDSS J003413.04$-$010026.9. 

\item 1RXS J092310.1$+$275448: A spectroscopically confirmed, $g=17.89\pm0.02$, $z=0.874\pm0.001$ QSO, SDSS J092314.20$+$275428.3, is coincident with the {\it Chandra}-detected source ($7.1\pm1.0 \times$\ctss) closest to the RASS position. This QSO is offset from the RASS position by $58$\asec.

In both of these cases, the separation between the RASS source positions and that of the SDSS counterparts to the probable {\it Chandra} re-detection is greater than 4$\times$ the quoted RASS positional uncertainty of 14\asec, so that the SDSS objects were not considered when these fields were selected as potentially hosting INSs. 

\item 1RXS J102659.6$+$364039: Another spectroscopically confirmed QSO, SDSS J102700.55$+$364016.0 ($g = 20.42\pm0.03$, $z=0.750\pm0.002$), is coincident with the only {\it Chandra}-detected source ($2.8\pm0.7 \times$\ctss) within this field, which is offset by $26$\asec\ from the RASS position. 

The spectrum of this SDSS object was taken post-DR4 and was therefore not available to us when we constructed our list of candidate INSs. However, its relatively small offset from the RASS position (the positional uncertainty is $12$\asec) and UV-excess ($(u-g) = 0.63\pm0.10$, $u=21.04\pm0.10$) mean that, according to the criteria described in \S~\ref{comp}, this field should not have been included in our list of candidate INSs. We note in addition that \cite{zickgraf03} proposed this object as one of two possible counterparts to this RASS source, although these authors did not classify it.

\item 1RXS J122344.6$+$373015: A faint SDSS-cataloged galaxy, SDSS J122344.96$+$373019.3 ($g = 22.13\pm0.07$), is coincident with the only {\it Chandra} source ($4.4\pm0.1 \times$\ctss) in this field, and offset $8$\asec\ from the RASS position. Its $(u-g) = 0.6\pm0.36$ suggests that this is a candidate QSO; its faintness in the $u$-band ($u = 22.74\pm0.36$) explains why its presence did not eliminate this field from consideration in Paper I. 
\end{itemize}

\section*{X-ray transients, spurious sources, or low-confidence detections}
In six cases our {\it Chandra} observations failed to detect an X-ray source within an arcminute of the RASS position even with a signal-to-noise threshold of 2. In most cases the offset to the nearest {\it Chandra} source is several arcminutes. At such large offsets it is very unlikely that these sources are the ones cataloged in the RASS, for which the positional uncertainties are generally $\sim$15\asec. In two other fields lowering the detection threshold does uncover a {\it Chandra}-detected source with a signal-to-noise ratio $<$3 offset by less than 30\asec\ from the RASS position. Optical spectroscopy (and potentially, deeper X-ray observations) would be required in both cases to confirm our tentative identifications of these X-ray sources based on the properties of their SDSS counterparts. 

\begin{itemize}
\item 1RXS J103415.1$+$435402: There are just two detected sources in the {\it Chandra} field; one is offset by over $3$\amin\ from the RASS position, while the other is nearly $4$\amin\ away. The fainter of the two has a count rate of $1.7\pm0.4 \times$\ctss, which is comparable to the {\it Chandra} count rate we predicted for 1RXS J103415.1$+$435402 ($1.8 \times$\ctss). This suggests that our observation was sensitive enough to detect this RASS source.

Lowering the threshold to 2 uncovers a source offset by 22.7\asec\ from the RASS position with a count rate of $0.9\pm0.3 \times$\ctss. This source is coincident with SDSS J103413.68$+$435345.5, a plausible candidate QSO ($(u-g) = -0.39\pm0.24$) that is too faint to have met the criteria for eliminating the field from consideration, as its $u=22.36\pm0.22$. 
 
\item 1RXS J110219.6$+$022836: The nearest {\it Chandra} source to the RASS position is offset by $8$\amin. This source has a count rate of $1.5\pm0.4 \times$\ctss, which is lower than our predicted {\it Chandra} count rate for the $\ROSAT$ source ($1.9 \times$\ctss). Again, our observation was sensitive enough to have detected this RASS source. 
Lowering the detection threshold does not uncover a {\it Chandra} source closer to the RASS position.

\item 1RXS J141944.5$+$113222: The nearest source to the RASS position is offset by $3.5$\amin. Its count rate ($1.6\pm0.4 \times$\ctss) is lower than the count rate we predicted for the $\ROSAT$ source ($2.4 \times$\ctss). 
Here again, lowering the detection threshold does not uncover sources closer to the RASS position.

\item 1RXS J142423.3$-$020201: No source is detected in this field with a signal-to-noise threshold of 3. Lowering the threshold to 2 does not uncover any detections offset by less than 10\amin\ from the RASS position.

\item 1RXS J151855.1$+$355543: With a detection threshold of 3, a single source is detected in this field; it is offset $19$\amin\ from the RASS position, and has a count rate of $2.9\pm0.9 \times$\ctss. This is comparable to our predicted {\it Chandra} count rate for the RASS source ($3.6 \times$\ctss) and suggests here again that our observation was sensitive enough to detect it. 

Lowering the threshold to 2 does uncover a source much closer to the RASS position (offset by 6.4\asec\ and with a {\it Chandra} count rate of $1.3\pm0.5 \times$\ctss), and which is positionally coincident with SDSS J151854.72$+$355538.4, a faint UV-excess object with a measured proper motion ($(u-g) = -0.40\pm 0.10$, $u = 21.25\pm0.08$,  $\mu = 0.25$ mas yr$^{-1}$) that should have eliminated this field from consideration in Paper I. Follow-up spectroscopy is required to uncover the nature of this object.

\item 1RXS J155705.0$+$383509: Here again, even lowering the detection threshold does not reveal any {\it Chandra} sources within 10\amin\ of the RASS position. 

\item 1RXS J162526.9$+$455750: The closest source to the RASS position is offset by $1.5\amin$. This source has a count rate of $4.5\pm0.5 \times$\ctss, above our predicted count rate for the $\ROSAT$ source ($1.5 \times$\ctss). However, the next closest source to the RASS position (offset by $4$\amin) has a count rate of $1.5\pm0.4\times$\ctss, comparable to this predicted {\it Chandra} count rate for the $\ROSAT$ source.
Lowering the threshold does not uncover detections closer to the RASS position.

\item 1RXS J205334.0$-$063617: Lowering the detection threshold still does not reveal any {\it Chandra} sources within 9\amin\ of the RASS position. 
\end{itemize}

\end{document}